\documentclass[%
reprint,
 amsmath,amssymb,
 aps,
]{revtex4-2}

\usepackage{graphicx}% Include figure files
\usepackage{dcolumn}% Align table columns on decimal point
\usepackage{bm}% bold math
\usepackage{nicefrac}
\usepackage{multirow,wrapfig,array,tabularx,rotating,verbatim,float,mathtools,booktabs,color}
\begin{document}

\title{Exploring organic semiconductors in solution: The effects of solvation,
alkylization, and doping}% Force line breaks with \\
%\thanks{A footnote to the article title}%

\author{Jannis Krumland}
\affiliation{%
 Humboldt-Universit{\"a}t zu Berlin, Physics Department and IRIS Adlershof, 12489 Berlin, Germany
}%
\affiliation{%
 Helmholtz-Zentrum Berlin, 12489, Berlin, Germany
}%

\author{Ana M. Valencia}%
\affiliation{%
 Humboldt-Universit{\"a}t zu Berlin, Physics Department and IRIS Adlershof, 12489 Berlin, Germany
}%
\affiliation{%
Carl von Ossietzky Universit\"at Oldenburg, Institute of Physics, 26129 Oldenburg, Germany
}%

\author{Caterina Cocchi}%
\affiliation{%
 Humboldt-Universit{\"a}t zu Berlin, Physics Department and IRIS Adlershof, 12489 Berlin, Germany
}%
\affiliation{%
Carl von Ossietzky Universit\"at Oldenburg, Institute of Physics, 26129 Oldenburg, Germany
}%

\date{\today}% 

\begin{abstract}
The first-principles simulation of the electronic structure of organic semiconductors in solution poses a number of challenges that are not trivial to address simultaneously.
In this work, we investigate the effects and the mutual interplay of solvation, alkylization, and doping on the structural, electronic, and optical properties of sexithiophene, a representative organic semiconductor molecule.
To this end, we employ (time-dependent) density functional theory in conjunction with the polarizable-continuum model. 
We find that the torsion between adjacent monomer units plays a key role, as it strongly influences the electronic structure of the molecule, including energy gap, ionization potential, and band widths.
Alkylization promotes delocalization of the molecular orbitals up to the first methyl unit, regardless of the chain length, leading to an overall shift of the energy levels.
The alterations in the electronic structure are reflected in the optical absorption, which is additionally affected by dynamical solute-solvent interactions. Taking all these effects into account, solvents decrease the optical gap by an amount that depends on its polarity, and concomitantly increase the oscillator strength of the first excitation.
The interaction with a dopant molecule promotes planarization. In such scenario, solvation and alkylization enhance charge transfer both in the ground state and in the excited state.
\end{abstract}

\maketitle

\section{Introduction}
Organic semiconductors are key components of advanced materials due to their efficiency in absorbing and emitting light~\cite{onei-kell11am,zhan+18am}. Their chemical versatility can be exploited to enhance these characteristics and, more generally, their physico-chemical properties, including their solubility and their ability to be doped, that can further tailor their functionalities~\cite{katz+01acr,anth06cr,usta+11acr,ruiz+12jpcl,jian+14acr}. Such great variability brings about a formidable structural and electronic complexity that is non-trivial to control experimentally nor to model theoretically. This is a challenging task especially for \textit{ab initio} quantum-mechanical simulations, where the chemical composition and the initial geometry of the system are the only input.
In spite of the proven success of first-principles methods in unraveling the electronic and optical properties of a variety of organic compounds or composite systems thereof~\cite{ruin+03sm,tiag+03prb,humm-ambr05prb,ortm+07prb,blas-atta11apl,zhu,shar+12prb,klet+16pccp,li+17prm,cocc+18pccp,ana1,ana2,schi+20jpcc}, the simplified models that are often adopted hinder the possibility to achieve a comprehensive picture, in which all (or at least most of) the involved degrees of freedom are taken into account. For example, assuming hydrogenated oligomers \textit{in vacuo} to describe polymers in solution not only fails to include solvation effects but also neglects the role of alkyl chains that enhance the solubility of extended segments~\cite{whyAlkylChainsAreAdded,mans+20jmcc,thom+18ml}. Moreover, \textit{ab initio} electronic structure calculations usually rely on force minimization procedures to obtain relaxed geometries which are ranked solely based on their energetics~\cite{wang+16jcp}. In this way, it is not uncommon to find local minima that are energetically very close to each other and that can be alternatively accessed depending on the initial conditions. 
For example, it is known that adjacent monomers in oligo- and polythiophenes can occur in an aligned (\textit{cis}) or an opposing orientation (\textit{trans}), representing two local minima.~\cite{viruela1997,raos2003} However, ordered thiophenes are usually associated with the \textit{trans} variant,~\cite{malavika,gao+13jmcc} which is energetically more favorable.~\cite{khla+16}

Further complexity is added by doping, which is routinely exploited in organic semiconductors to improve their electronic performance~\cite{walzer2007cr,yim+08am,luessem2013pssa,mendez2013,gao+13jmcc,pingel2013,salz+16acr,jaco-moul17am}. The interaction with dopant species leads to supramolecular compounds with unique characteristics that are determined by the hybridization between the donor and the acceptor and by the charge transfer between them~\cite{baku+12sci,broc+13jpcc,mendez2013,mendez2015,salz+16acr,beye+19cm,belo+20jpcc}. Also in this scenario, \textit{ab initio} simulations \textit{in vacuo} have successfully led to the understanding of the basic mechanisms driving the electronic and optical activity of these complexes~\cite{wing-bred11jpcc,zhu,mendez2015,beye+19cm,ana1,ana2}. However, open questions concerning the role of solvents and functional chains have not found an answer yet.

In this paper, we present a systematic analysis of the effects of solvation, functionalization, and doping of organic semiconductors in solution, carried out within the quantum-mechanical \textit{ab initio} framework of time-dependent density-functional theory. We examine how these parameters and their interplay influence structural, electronic, and optical properties and, in turn, how these characteristics affect each other. For this purpose, we consider a sexithiophene oligomer in the \textit{trans} conformation, a representative organic semiconductor, and investigate how the presence of alkyl chains, the interaction with solvents of increasing polarity, as well as $p$-doping perturbs its intrinsic properties. We discuss which initial conditions have to be fulfilled to capture these effects in the adopted theoretical scheme, and which degrees of freedom influence the accuracy of the results.

%%%%%%%%%%%%%%%%%%%%%%%%%%%%%%%%%%%%%

\section{Theoretical Background and Computational Details}
The electronic properties of the molecules investigated in this work are determined using density functional theory\cite{hohenbergKohn1964} (DFT) within the generalized Kohn-Sham (KS) framework\cite{ks1965,generalizedKS1}. In this formalism, the nonlinear single-particle KS equation,
\begin{align}
\hat{{\cal H}}_{\text{KS}}[\rho]\psi = E\psi,
\end{align}
where the electron density is defined as 
\begin{align}
    \rho=\sum_k^{\text{occ}}|\psi_k|^2,
\end{align}
is solved iteratively until self-consistency is achieved, yielding the KS orbitals, $\psi_k$, and the corresponding KS eigenvalues $E_k$. In atomic units, %($\hbar$~=~$a_0$~=~$e$~=~$m_e$~=~1)
the electronic KS Hamiltonian is given by
\begin{align}\label{ks.eq}
\hat{\cal H}_{\text{KS}}[\rho](\textbf{r}) = -\frac{\nabla^2}{2}+v_{\text{nuc}}(\textbf{r})+v_{\text{R}}[\rho](\textbf{r}) + v_\text{H}[\rho](\textbf{r}) + {v}_{\text{xc}}[\rho](\textbf{r}),
\end{align}
where $v_{\text{nuc}}$ is the electrostatic potential due to the nuclei, $v_{\text{R}}$ is the reaction potential caused by surrounding solvent molecules (see below), $v_{\text{H}}$ is the Hartree potential, \textit{i.e.} the electrostatic potential created by the electron density $\rho$, and ${v}_{\text{xc}}$ comprises the residual quantum-mechanical aspects of the electron-electron interactions. The last term must be approximated, as its exact form is unknown. In the original Kohn-Sham theory, an explicitly density-dependent local potential is assumed~\cite{ks1965}. Thanks to its excellent trade-off between computational efficiency and overall accuracy, this recipe was widely adopted for decades. However, for almost thirty years, it has proven of value, especially in the context of (doped) organic materials, to give up on this restriction of locality and to mix in a portion of exact exchange, calculated from the Hartree-Fock theory~\cite{b3lyp}. For these \textit{hybrid functionals}, ${v}_{\text{xc}}$ is no longer a universal local potential, but instead a non-local operator, depending on the density $\rho$ as well as on the set of (occupied) orbitals $\{\psi_k\}$~\cite{kuemmel2008rmp}. For organic materials, hybrid functionals usually yield more accurate electronic and optical properties in comparison to their (semi)local counterparts~\cite{sousa2007jpca,sini+11jctc}. Going one step further, great popularity has been gained in the last decade by \textit{range-separated} hybrid functionals~\cite{atal+16prb,gall+16jctc}, in which the Coulomb interaction entering the exact-exchange energy is separated into short- and long-range parts\cite{camb3lyp}:
\begin{align}
    \frac{1}{r} = \frac{\alpha + \beta \,\text{erf}({\mu r})}{r} + \frac{1-[\alpha + \beta \,\text{erf}({\mu r})]}{r},
\end{align}
where $\alpha$, $\beta$, and $\mu$ are adjustable parameters. In this way, the exact-exchange energy becomes the sum of two terms, the first of which, dominant in the short range, is replaced by an approximate DFT functional. In contrast to their semilocal and global hybrid counterparts, these functionals are able to correctly describe excited states involving significant charge transfer~\cite{maitra2017, ploetner+2010jctc}.

To account for solvation effects, we employ the integral-equation formalism~\cite{cances1997jcp} of the polarizable continuum model~\cite{tomasi2005cr}. In this framework, the solvent is modelled as a bulk dielectric enclosing the solute molecule. The solvent is characterized by its dielectric constant~$\epsilon$. The charge densities associated with the electrons and nuclei of the molecule polarize the continuum solvent, creating interfacial polarization charges that generate the reaction potential $v_{\text{R}}[\rho](\textbf{r})$ acting back onto the molecule. The quantum-mechanical equations for the molecule are solved self-consistently in combination with the electrostatic ones for the environment.
The adopted approach neglects dispersion and chemical interactions between solute and solvent, which can become significant in some cases. To capture these effects, more sophisticated modelling would be in order, entailing increased computational complexity. However, for the scope of this work, which is determining polarity-dependent trends, it is sufficient to account for electrostatic interactions only.

Optical absorption spectra are calculated within linear-response time-dependent DFT (TDDFT)~\cite{rungeGross1984}, which implies solving the matrix equation
\begin{align}
    \begin{pmatrix}
    \textbf{A} & \textbf{B}\\
    \textbf{B}^* & \textbf{A}^*\\
    \end{pmatrix}
    \begin{pmatrix}
    \textbf{X}\\
    \textbf{Y}
    \end{pmatrix}
    =\omega
    \begin{pmatrix}
    \textbf{1} & \textbf{0}\\
    \textbf{0} & \textbf{-1}\\
    \end{pmatrix}
    \begin{pmatrix}
    \textbf{X}\\
    \textbf{Y}
    \end{pmatrix}\label{casida.eq}
\end{align}
for the excitation energies $\omega_k$ and (de-)excitation coefficients $\textbf{X}_k$ ($\textbf{Y}_k$). For hybrid functionals, the coupling matrices $\textbf{A}$ and $\textbf{B}$ are suitable interpolations\cite{chiba2006jcp} between those from TDDFT and time-dependent Hartree-Fock theory~\cite{casida}.

 For time-dependent calculations including a solvent, a term corresponding to $v_{\text{R}}$ is added to the matrix elements of $\textbf{A}$ and $\textbf{B}$, accounting for the solvent polarization due to the excited non-stationary electron density\cite{cossi2001jcp, corniMetalSurface}. These excitation dynamics are so fast that the solvent molecules cannot adapt to it by reorienting themselves, and thus they are merely polarized. This effect is taken into account in non-equilibrium solvation models, in which the solvent is characterized not only by its static dielectric constant $\epsilon$, but also its high-frequency limit, $\epsilon_\infty$. The latter is connected to the usual refractive index $n$ via the well-known relation $\epsilon_\infty=n^2$. The static value $\epsilon$ corresponds to full solvent equilibration to the solute state, whereas $\epsilon_\infty$, describes the mere electronic polarization. Hence, ground-state calculations assume a surrounding continuum environment with dielectric constant $\epsilon$, whereas the time-dependent excitation-induced charge density interacts with an effective solvent of dielectric constant $\epsilon_\infty$.
 
 In addition to linear-response TDDFT for solvated systems, we make use of the complementary state-specific approach\cite{improta2006jcp}. In this framework, the stationary excited-state electron density and the corresponding solvent polarization are determined self-consistently. 
 This approach is different from the linear-response method described above, in which absorption spectra are the main output. 
 Within the state-specific approach, excitation energies are obtained as $\omega = {\cal G}^{\mathrm{neq}}_e - {\cal G}_g$, where ${\cal G}_g$ and ${\cal G}^{\mathrm{neq}}_e$ are the ground-state and the non-equilibrium excited-state free energies resulting from a ground-state and a state-specific-TDDFT calculation  respectively. In the latter, only the fast degrees of freedom of the solvent are allowed to adapt to the variation of electron density (non-equilibrium solvation). We will explore the complementary nature of the linear-response and state-specific approaches by applying them to different types of excitations.

 We make use of the natural population analysis~\cite{npa}, which has been devised to correct the major deficiencies of the well-known Mulliken charge analysis~\cite{mulliken1955jcp}. In the spirit of other ``natural orbital" methods (\textit{e.g.}, natural transition orbitals in the context of optical excitations\cite{martin2003jcp}), the idea behind the natural population analysis is the construction of minimal sets of atomic orbitals harboring the majority of the molecular electronic density around those atoms. 
 
 %%%%%%%%%%%%%%%%%%%%%%%%%%%%%%%%%%%%%%%

%\section{Computational Details}

% 0.19, 0.46, and 0.33 a, b, mu

All calculations are performed using the \textsc{Gaussian}16 package\cite{gaussian}. The CAM-B3LYP range-separated hybrid functional~\cite{camb3lyp, camb3lyp1, camb3lyp2} is employed in conjunction with the double-$\zeta$ cc-pVDZ and the triple-$\zeta$ cc-pVTZ basis sets~\cite{basisSets, elephant}. 
We employ the polarized double-zeta (double-$\zeta$) basis set, cc-pVDZ, for geometry optimizations and TDDFT calculations. For electronic properties such as orbital energies and partial charges, we use the triple-$\zeta$ pendant, cc-pVTZ. 
This choice enables quantitative comparisons to previous results obtained for related materials~\cite{ana1,ana2,malavika}. We checked that cc-pVDZ gives almost identical results to the 6-31G(d,p) basis set~\cite{popleBasisSets} in TDDFT calculations. For the excited-state calculations, we also considered the effect of adding diffuse functions, using the 6-31+G(d,p) and 6-31++G(d,p) basis sets, finding only minor improvements in accuracy (see Section~\ref{optical_properties_ctc.sec} and S4.5 in the Supplementary Material).
Similar to the ground-state case, also the self-consistent excited-state density obtained from state-specific TDDFT can be used to analyze partial charges. Although this excited-state analysis is not as commonly employed as in the ground state, it contributes to understand the character of the excitations. For a quantitative comparison with the partial charges computed from the ground-state density, we use the triple-$\zeta$ basis set also for the corresponding state-specific TDDFT calculations.

Long-range dispersion interactions are included in the geometry optimizations through Grimme's empirical correction scheme with the original D3 damping function.~\cite{grimme1}
Having checked that these contributions play no role in the TDDFT results, we did not include them in those calculations. 
As implicit solvents, we consider benzene ($\epsilon$~=~2.27, $n$~=~1.50), chloroform ($\epsilon$~=~4.71, $n$~=~1.45), and nitromethane ($\epsilon$~=~36.5, $n$~=~1.38), representing apolar, semipolar, and polar solvents, respectively. We emphasize once more that the purely electrostatic treatment of the PCM is not necessarily adequate for these particular solvents;~\cite{menucci+2002jpca} they are chosen solely based on the values of their dielectric constants.

%%%%%%%%%%%%%%%%%%%%%%%%%%

\section{Results}
\subsection{Structural and Electronic Properties}
\subsubsection{\textit{In Vacuo}}

\begin{figure}
    \centering
    \includegraphics[width=0.48\textwidth]{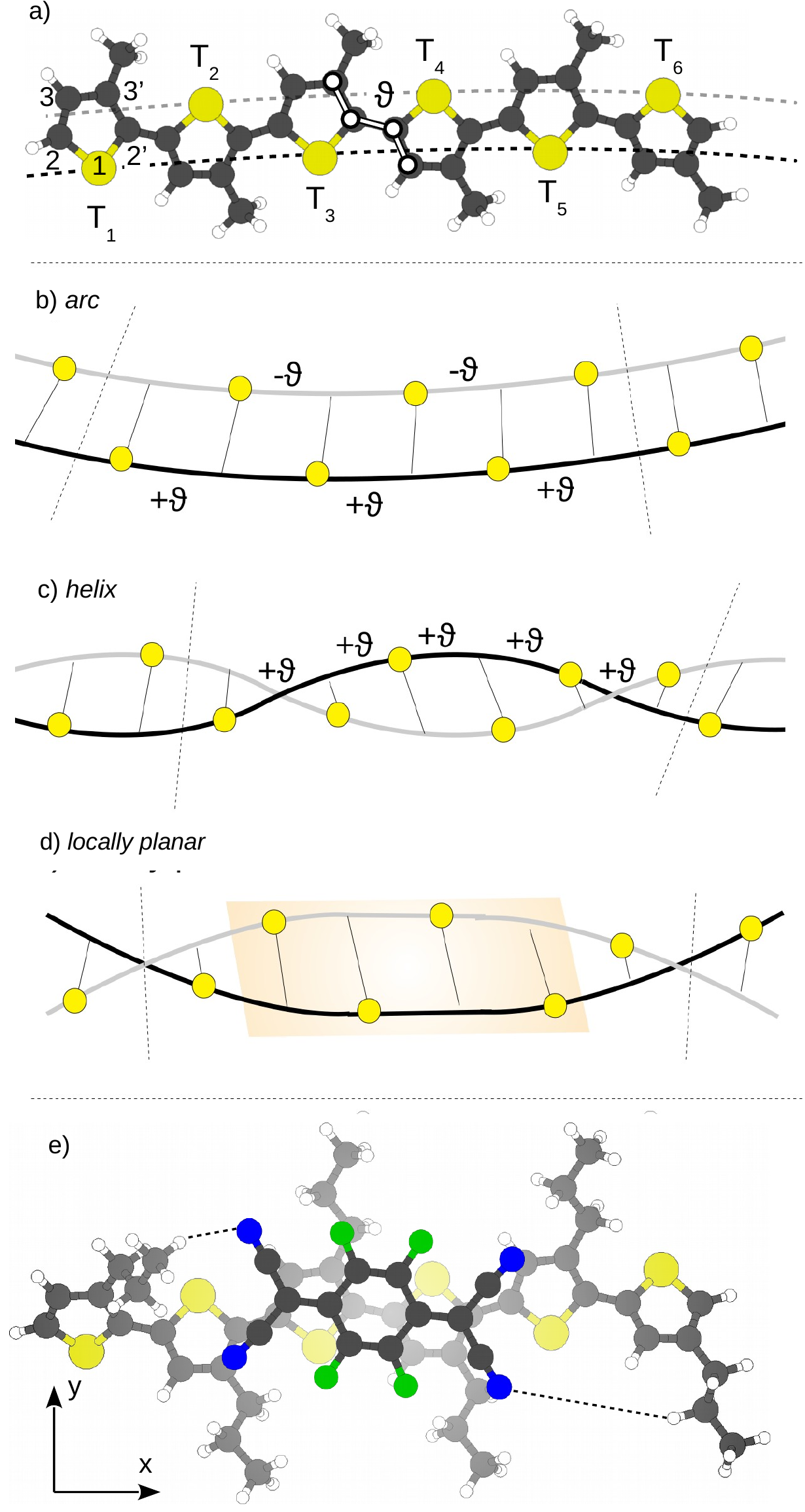}
    \vspace{0.5cm}
    \caption{a) Ball-and-stick representation of 6T, with the torsion angle $\theta$ and the alkyl chain docking sites indicated. In the first ring (T$_1$), the intra-monomeric numbering of the atoms is shown. b-d) Schematic representation of polythiophene segments corresponding to different 6T conformations. Yellow circles mark the positions of the S atoms. e) Equilibrium geometry of the propylated 6T/F4-TCNQ charge-transfer complex.}
    \label{geometries.fig}
\end{figure}

We examine sexithiophene (6T) as a representative organic semiconductor, which is often adopted to model extended polymeric poly(3-hexylthiophene) (P3HT) chains~\cite{piConjugation,gierschner2007,yin+16jpcc,nigh+18jpcc}. 
Like all oligothiophenes, this molecule is polymorphic both in the gas phase~\cite{pan+07am} and in its crystalline form~\cite{serv+94cm,anto+98am,herm+05jpca,pith+15crd,klet+16pccp}.
Here, we consider 6T in the \textit{trans} conformation, in which the S atoms in consecutive thiophene units (T) point to opposite directions [see Figure \ref{geometries.fig}a)].
This conformation represents the global energetic minimum,~\cite{viruela1997,raos2003} and is most commonly studied also in the context of extended thiophene polymers~\cite{gao+13jmcc,khla+16,malavika} due to its enhanced order compared to the \textit{cis} conformer.

It is known that 6T is not planar, neither \textit{in vacuo} nor in solution. Neighboring T units are twisted with respect to each other, striking a balance between the planarizing $\pi$-conjugation and the steric S$\cdots$H repulsion, which is partially relieved by torsion of the rings\cite{viruela1997}. The torsion equips \textit{trans} 6T with a large number of sub-conformations. We find that adopting a planar geometry as a starting point for the structure optimization of 6T typically leads to an unstable planar conformation or to an irregular structure, in which the individual rings are bent up- or downwards without a pattern. 
The planar arrangement of two neighboring T units corresponds to a saddle point of the torsional potential energy curve~\cite{orti1195,raos2003,viruela1997}; numerical noise tips the scales in either direction at some point of the calculation, giving rise to a random sign sequence of the torsion angles across the overarching structure. This scenario is common when the geometry optimization starts from a planar structure, but ends in a non-planar one. 

\begin{figure}
    \centering
    \includegraphics[width=0.50\textwidth]{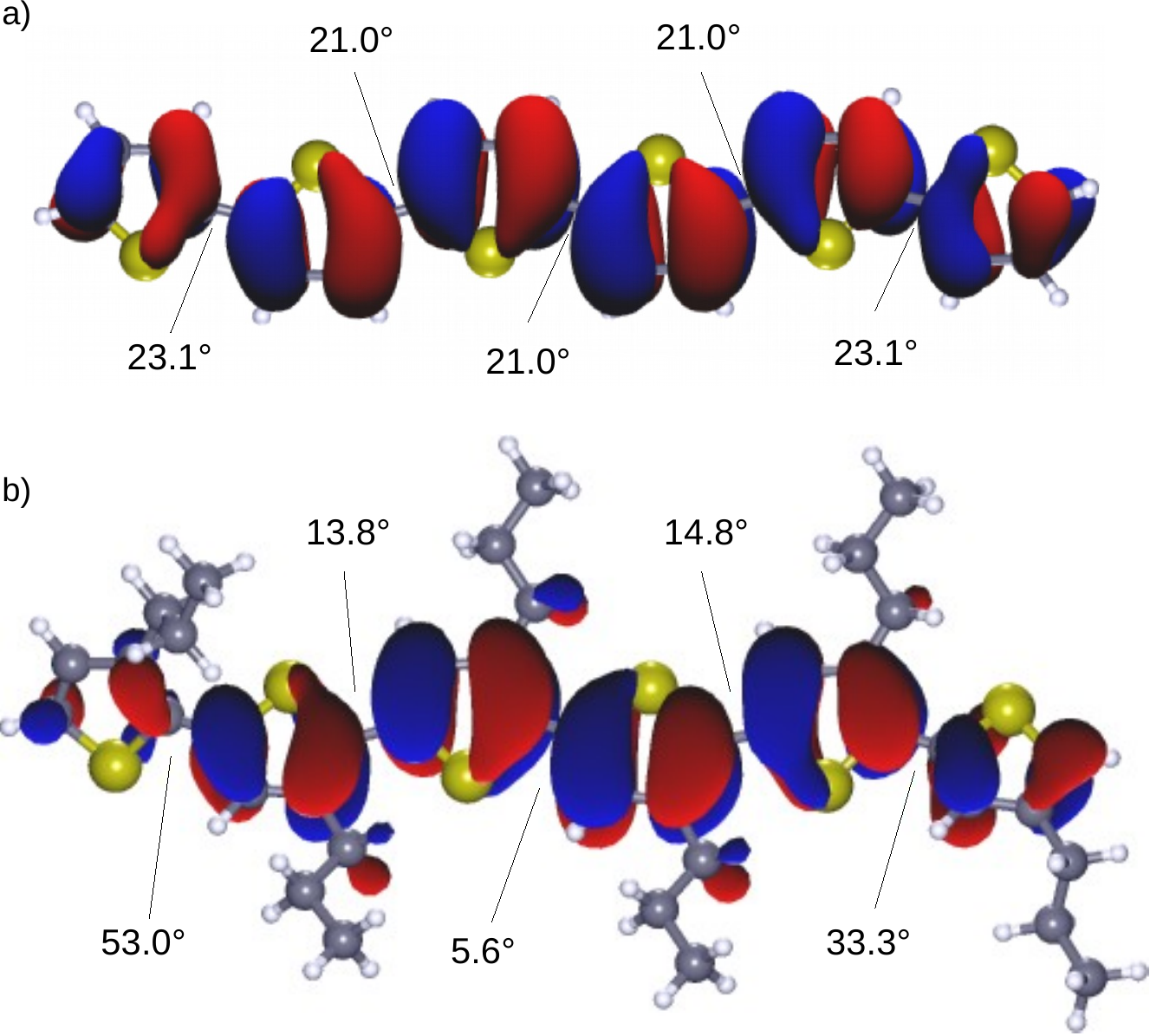}
    \caption{Isosurfaces of the HOMO of 6T \textit{in vacuo}, a) in the \textit{arc} configuration without alkyl chains, and b) in the geometry extracted from the CT complex (\textit{locally planar}), with C$_3$H$_7$ groups attached. The torsion angles between adjacent T rings are specified. The isovalue for the orbitals is fixed at $\pm$0.016~$\mathrm{bohr}^{-3/2}$.}
    \label{homos.fig}
\end{figure}

To remove this arbitrariness and avoid unrealistic geometries corresponding to unstable saddle points, calculations can be nudged towards a more regular structure by assigning small starting torsion angles. This way, we obtain two geometries, which can be considered as opposite poles comprising all the intermediate irregular configurations in between. In one extreme, the torsion angle alternates from $+\theta$ to -$\theta$ from one T-T link to the next. This gives rise to the structure sketched in Figure~\ref{geometries.fig}b), characterized by an \textit{arc}-shaped backbone. In the other extreme, the torsion angle between subsequent T units maintains a fixed sign, resulting in the \textit{helical} structure shown in Figure~\ref{geometries.fig}c). The total energy difference between the two conformations is almost negligible, amounting to about 10~meV. The analysis of their vibrational frequencies, which are all positive, confirms that they correspond to energetic minima. In all cases, the average torsion angle is about 20$^\circ$. The angles are slightly larger for the outer rings, T$_1$ and T$_6$ [Figure \ref{homos.fig}a)], indicating that twists are especially favorable at the edge. In the following, we will consider these two limiting configurations in parallel, in order to achieve a confidence interval that encompasses all intermediate structures.

Before proceeding along these lines, we briefly discuss the choice of the exchange-correlation functional in the structural optimization. While semi-local generalized-gradient approximations such as PBE~\cite{pbe} are very efficient for obtaining reasonable equilibrium geometries, they are unsuitable for the considered systems, as they generally predict a flat structure for 6T. This failure is partially cured by employing a global hybrid functional like B3LYP\cite{b3lyp}. However, comparisons with higher-level quantum-chemical methods, such as M{\o}ller-Plesset perturbation theory~\cite{mp} and coupled cluster~\cite{ccsd}, reveal that global hybrid functionals still overstabilize the planar configuration and overestimate the torsional barrier~\cite{karpfen1997,choi1997}. On the other hand, range-separated hybrids like CAM-B3LYP~\cite{camb3lyp}, while mainly devised to improve the description of charge-transfer excitations~\cite{maitra2017}, appear to perform better also for the structural properties of $\pi$-conjugated oligomers\cite{LinAndLin, bla1, bla2}.

In order to increase their solubility, oligo- and polythiophenes are usually functionalized with alkyl side chains~\cite{whyAlkylChainsAreAdded, malavika, mans+20jmcc, thom+18ml}. These groups exert an influence on the structural and electronic properties of the backbone, which we explore in the following. The \textit{arc} and \textit{helix} configurations of 6T remain energetically equivalent upon replacing the H atoms highlighted in Figure~\ref{geometries.fig}a) by methyl (CH$_3$), ethyl (C$_2$H$_5$), or propyl (C$_3$H$_7$) groups. Their addition leads to an increase of the torsion angle $\theta$  [Table \ref{torsion.tbl}a)] due to the steric repulsion between the attached C atom of the alkyl chain and the close-by S atom\cite{methylTwist1, methylTwist2, methylTwist3}, which is stronger than the S$\cdots$H steric interaction, as a consequence of the higher van-der-Waals radius of C compared to the one of H. Since the innermost C atom of the chain is the crucial one in this context, replacing the H atoms by CH$_3$ groups causes an increase of 8-9$^\circ$ in the torsional angle, while extending the chains to C$_2$H$_5$ and C$_3$H$_7$ only causes slight further changes (1-2$^\circ$). 

\begin{table}[]
\caption{Torsion angle (in $^\circ$) between T$_3$ and T$_4$ for 6T in the \textit{arc}, \textit{helix}, and \textit{locally planar} configurations a) \textit{in vacuo} with different groups attached and b) In different solvents with CH$_3$ groups attached}
    \begin{flushleft}
    \footnotesize{a)}
    \end{flushleft}
    \centering
\begin{tabular}{p{1.4cm}p{1.2cm}p{1.2cm}p{1.2cm}}
\hline
     \centering\textbf{group} &  \centering \textit{arc}  & \centering \textit{helix} & \centering \textit{planar} \arraybackslash\\
     \hline \centering none & \centering 21.0 & \centering 17.5 & \centering 10.3\arraybackslash\\
     \centering CH$_3$ & \centering 28.6 & \centering 26.9 & \centering 8.2\arraybackslash\\
     \centering C$_2$H$_5$ & \centering 30.0 & \centering 28.4 & \centering 7.6\arraybackslash\\
     \centering C$_3$H$_7$ & \centering 30.7 & \centering 28.9 & \centering 5.6\arraybackslash\\
     \hline 
\end{tabular}
    \begin{flushleft}
    \footnotesize{b)}
    \end{flushleft}
\begin{tabular}{p{1.4cm}p{1.2cm}p{1.2cm}p{1.2cm}}
\hline
    \centering \textbf{solvent} & \centering \textit{arc} & \centering \textit{helix} & \centering \textit{planar} \arraybackslash\\
     \hline \centering none  & \centering 28.6 & \centering 26.9 & \centering 8.2\arraybackslash\\
      \centering C$_6$H$_6$  & \centering 25.6 & \centering 21.8 & \centering 6.5\arraybackslash\\
     \centering CHCl$_3$  & \centering 23.7 & \centering 19.9 & \centering 5.8\arraybackslash\\
     \centering CH$_3$NO$_2$  & \centering 20.8 & \centering 12.8 & \centering 5.0\arraybackslash\\
     \hline
\end{tabular}
\label{torsion.tbl}
\end{table}

\begin{figure}
    \centering
    \includegraphics[width=0.47\textwidth]{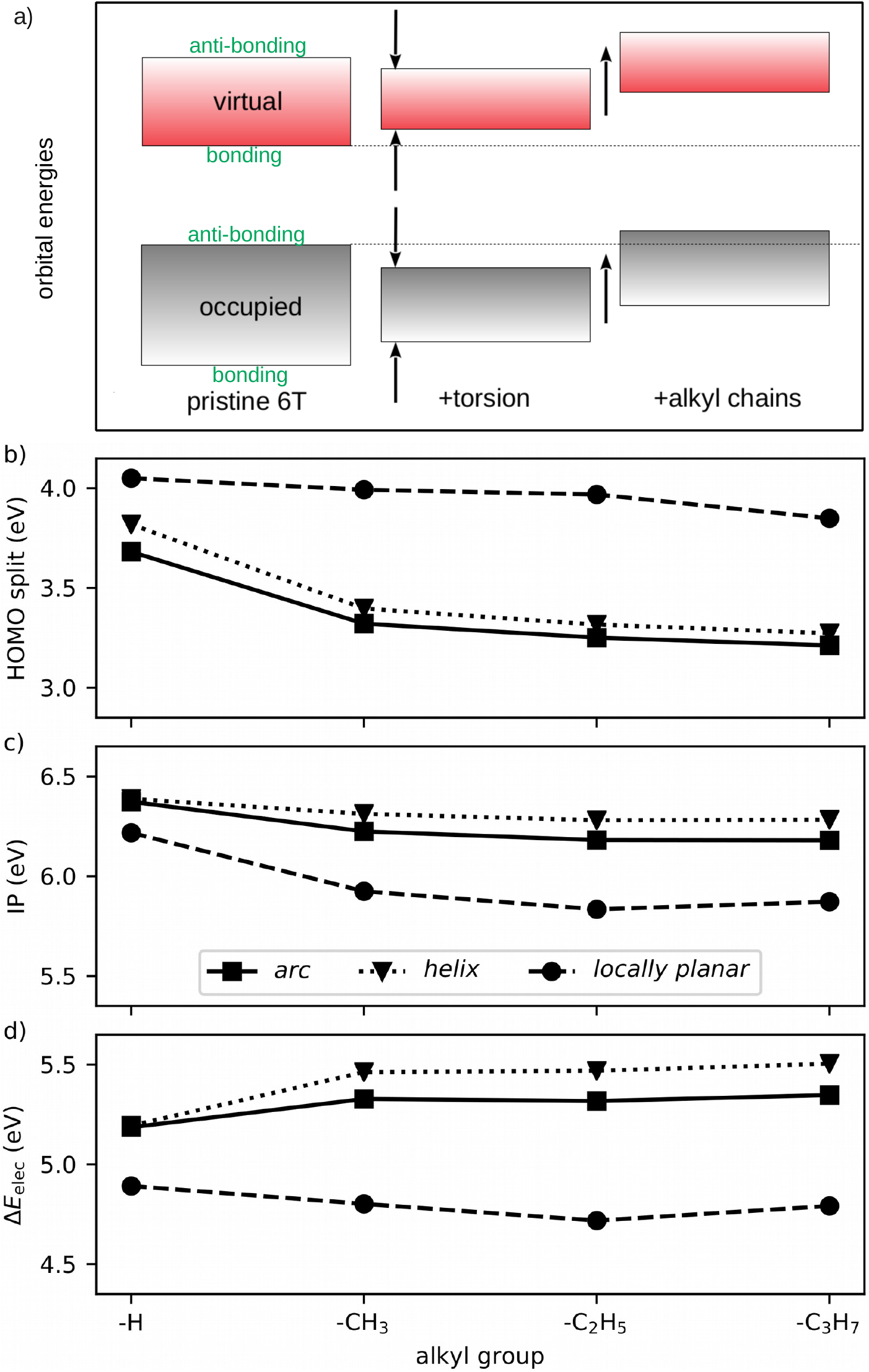}
    \vspace{0.5cm}
    \caption{a) Schematic representation of the indirect torsion-related and direct effects of alkyl chains on the valence and conduction bands of 6T.
    b) HOMO (H) splitting ($E_{\text{HOMO}}-E_{\text{HOMO-11}}$),  c) ionization potential $-E_{\text{HOMO}}$, and d) energy gap, $\Delta E_{\text{elec}}$, computed as $E_{\text{LUMO}}-E_{\text{HOMO}}$, of 6T in the \textit{arc}, \textit{helix} and \textit{locally planar} configurations, with different alkyl chains.}
    \label{levels.fig}
\end{figure}

The alkyl groups affect the electronic properties of the 6T in two different ways [Fig. \ref{levels.fig}a)]. The first one is related to the aforementioned variability of the torsion angles. These angles are key degrees of freedom, as they can be manipulated at very small energetic costs and concomitantly exert great influence on the electronic structure of the molecule~\cite{zade2007}, which is a general feature of flexible organic semiconductors~\cite{gierschner2007}. The highest occupied molecular orbital (HOMO) of 6T is the fully antibonding superposition of the six HOMOs of the individual T rings. The corresponding fully bonding superposition is not the HOMO-5, as one might expect, but the HOMO-11, as the HOMO-1 of the T units form an intermediate dispersionless band~\cite{tightBinding} (see Section~S1 in the Supplementary Material for detailed information about the frontier energy levels). The splitting between the KS energies of the HOMO and the HOMO-11 ($E_{\text{HOMO}}-E_{\text{HOMO-11}}$), which can be seen as the finite equivalent of the valence bandwidth, gives an estimate of the electronic coupling between the rings~\cite{bredas2004}. From the results shown in Figure~\ref{levels.fig}b), it is evident that this quantity decreases dramatically in the presence of covalently bonded alkyl groups to the \textit{arc} and \textit{helix} 6T. This energy variation is particularly pronounced between H-terminated 6T and its CH$_3$-functionalized counterpart. Prolonging the chains to C$_2$H$_5$ and C$_3$H$_7$ induces only slighter decrements, mirroring the trend of the torsion angles [Table \ref{torsion.tbl}a)]. The correlation between these two parameters indicates that increasing torsion is indeed responsible for the change of the valence electronic structure. Additional influence of the alkyl chains is evident upon inspection of the ionization potential (IP). This quantity, estimated as $-E_{\text{HOMO}}$, according to the Koopmans' theorem for DFT~\cite{koopmans}, \textit{decreases} when alkyl groups are attached to the 6T [Figure~\ref{levels.fig}c)]~\cite{cocchiGNF}. This effect cannot be understood in terms of the torsion-induced decoupling of the T units, which conversely implies a decrease of $E_{\text{HOMO}}$, \textit{i.e.}, an \textit{increase} of the IP: since the HOMO of 6T is the fully antibonding linear combination of the HOMOs of the T units, its energy is lowered when the coupling is reduced. The reduction of the IP upon alkyl functionalization is instead related to the partial delocalization of the frontier orbitals, which extend into the covalently bonded groups. However, as shown in Figure~\ref{homos.fig}b) for the HOMO, this spill-out charge does not extend beyond the first unit of the alkyl chain. 

The relation between torsion angle and low-lying virtual orbitals is essentially reversed with respect to the scenario delineated above for the occupied states. Decreasing the coupling between the rings by increasing the torsion raises the energy of the lowest unoccupied moleular orbital (LUMO), as it corresponds to the fully bonding superposition of the LUMOs of the T units\cite{tightBinding}. This increase adds up to the influence of the alkyl chains, which increase the energy of all levels (occupied and virtual), resulting in a significantly expanded electronic gap $\Delta E_{\text{elec}}=E_{\text{LUMO}}-E_{\text{HOMO}}$ [Figure~\ref{levels.fig}d)]. \textit{Arc} and \textit{helix} configurations differ significantly with respect to each other in terms of $\Delta E_{\text{elec}}$, due to differences between their respective torsion angles, which, while small, entail sizeable variations of the electronic levels.
Since the main change in torsion is observed upon replacing H by CH$_3$, and since the frontier orbitals do no extend further than on the first unit of the chains, the effect of covalently attached alkyl groups on the electronic structure of 6T is essentially captured by the addition of CH$_3$ groups (see also Section~S1 of the Supplementary Material).

%%%%%%%%%%%%%%%%%%%%
\subsubsection{In solution}\label{electronic_properties_6t_sol.sec}
For the analysis of the structural properties of 6T in solution, we focus only on CH$_3$-functionalized 6T. We choose this variant over the H-terminated one, since solvated P3HT is usually alkylized.
We examine three scenarios corresponding to three solvents with increasing polarity: benzene (C$_6$H$_6$), which is an \textit{apolar} solvent, chloroform (CHCl$_3$), which can be considered a \textit{semipolar} solvent, and the strongly \textit{polar} nitromethane (CH$_3$NO$_2$). Assuming purely electrostatic coupling, apolar solvents interact with the solute through (local) dipole$\cdots$induced dipole interactions, whereas polar ones couple by stronger dipole$\cdots$dipole forces. Computationally, they are treated on equal footing.

The presence of a solvent reverses the increase of torsion angles upon alkyl functionalization [Table \ref{torsion.tbl}b)]. Torsion angles decrease monotonically with the polarity of the solvent; the smallest angles are thus found in 6T in CH$_3$NO$_2$, the \textit{polar} solvent. Sizeable differences in this reduction exist between \textit{arc} and \textit{helix} conformations: In the former, the torsion is reduced by approximately 30\%, while in the latter by about 50\%, when comparing the corresponding values \textit{in vacuo} and in CH$_3$NO$_2$. The angle of 12.8$^\circ$ assumed by \textit{helix}-6T in the \textit{polar} solvent corresponds to an almost planar structure. We conjecture that the underlying cause for the planarization is the direct electrostatic solute-solvent interaction. While the solvent also causes a redistribution of electronic charge within the solute that potentially gives rise to planarizing forces on the nuclei, we find this redistribution to be so small that this effect is likely negligible. This is supported by the fact that the bond lengths along the conjugation path remain essentially unaffected by solvation, indicating that the electronic structure of the conjugated network is not perturbed by the solvent\cite{piConjugation}. The S atom and the alkyl chain exchange a very small fraction of charge upon solvation ($\sim$0.01~$e$), and the local dipole moments of neighboring T units remain so small that inter-ring dipole-dipole forces are weak and likely negligible here.

It is tempting to analyze the electronic structure of the solvated molecule in terms of KS eigenvalues, as we did for the systems \textit{in vacuo}. However, this type of analysis is based on the assumption that the KS energies represent reasonable estimates for IPs and electron affinities, \textit{i.e.}, that they approximate well the energy required to add or remove an electron to the system. The resulting charged systems evoke strong solvent responses: With a charged solute, the polarization charge densities at the interface between solute and solvent carry a total charge with opposite sign~\cite{cances1997jcp}, which is not the case for neutral species. The stabilizing solute-solvent interaction is thus of monopol-monopol type and much stronger than the dominant dipole-dipole interactions encountered in neutral solutes. KS eigenvalues, however, implicitly assume a frozen solvent. Thus, they represent the unphysical situation in which the charged solute interacts with the reaction field of the neutral species, grossly underestimating the degree of stabilization. Indeed, KS electronic gaps remain nearly unchanged, while the stabilization should lead to a sizeable bandgap decrease on the order of $\sim$1~eV~\cite{neaton2006prl}. In vertical ionization processes, the solvent does not instantaneously equilibrate in full with the freshly ionized species; only the fast degrees of freedom are able to do so. Thus, vertical ionization energies should be calculated with a total free energy-based non-equilibrium solvation approach similar to the one commonly employed for optical excitations\cite{ipNonEq1, ipNonEq2}. We do not follow this path here. The strong level renormalization is not mirrored in the optical spectra, since optical excitations are charge-neutral; shifts in excitation energies are on the order of 100~meV, as we will see in Section~\ref{optical_properties.sec}. The decrease of the electronic bandgap is, to a large extent, compensated by a corresponding decrease of the exciton binding energy due to the screening effect of the continuum solvent.

%%%%%%%%%%%%%%%%%%%%%%
\subsubsection{6T/F4-TCNQ charge-transfer complex}
\label{electronic_properties_ctc.sec}
The electronic structure and in particular the conductivity of organic semiconductors can be effectively enhanced by molecular doping.
In oligo- and polythiophenes, the strong electron acceptor 2,3,5,6-tetrafluoro-7,7,8,8-tetracyanoquinodimethane (F4-TCNQ) has shown to be a particularly efficient $p$-dopant~\cite{yim+08am,mendez2015,li+16oe,jacobs2016,hynynen2017mm,hamidi2017afm}. 
% BEFORE
%However, numerous studies have shown that such interactions are far from trivial~\cite{malavika, pingel2013, gao2013, degradingICT, jacobs2018, neelamraju2018}. Both oligothiophenes and P3HT form charge transfer (CT) complexes with F4-TCNQ, although in P3HT integer CT can occur as well~\cite{salz+16acr,wang+15prb,mendez2015,malavika}.
% AFTER
However, numerous studies have shown that such interactions are far from trivial; both oligothiophenes and P3HT form charge transfer (CT) complexes with F4-TCNQ, although in the polymer, integer CT can occur as well~\cite{malavika, pingel2013, gao2013, degradingICT, jacobs2018, neelamraju2018, salz+16acr,wang+15prb,mendez2015,localizedCT}.
The prevalence of either form of CT is mainly determined by the degree of order in the P3HT~\cite{pingel2013}. Highly-ordered regioregular P3HT has a tendency to aggregate and thus form large planar domains, enabling polaron delocalization and charge separation~\cite{gao2013}. However, there is increasing evidence for the coexistence of both types of CT also in the ordered thiophene polymer, with the relative occurrence depending on doping concentration and processing parameters~\cite{degradingICT, jacobs2018, neelamraju2018}. Disordered polymers, like the regiorandom variant of P3HT, aggregate and planarize to a much smaller degree than ordered polymers. Without extended planar regions, CT complex formation prevails over ion pair formation~\cite{malavika}. While integer CT enhances the conductivity of polymers much more than partial CT, oligomeric crystals as well exhibit a significant increase in conductivity upon F4-TCNQ admixture, in spite of the exclusive formation of CT complexes~\cite{mendez2015}.

We simulate CT complexes by combining 6T with F4-TCNQ molecules to form $\pi$-$\pi$ stacked structures. Regardless of the presence of alkyl chains bound to the 6T, in the optimized geometry of the complex, the tetrafluorobenzene ring of the acceptor is situated above the link between T$_3$ and T$_4$~\cite{zhu, ana1, ana2} [see Figure~\ref{geometries.fig}e)]. The four T rings underneath the F4-TCNQ are almost flat. On the other hand, the outermost rings, T$_1$ and T$_6$, are bent upwards. We term this 6T configuration as \textit{locally planar}, bearing in mind that it exists only as part of the CT complex. Without alkyl chains attached and \textit{in vacuo}, the torsion angle of the outer rings amounts to 23.9$^\circ$. Inclusion of CH$_3$ groups increases this value to 34.0$^\circ$ on average. 
Prolonging the chain length increases the torsion, from 36.9$^\circ$ with C$_2$H$_5$ up to 41.9$^\circ$ with C$_3$H$_7$ groups. Notably, the last value (41.9$^\circ$) is the average between 53.0$^\circ$, obtained between T$_1$ and T$_2$, and 33.3$^\circ$, between T$_5$ and T$_6$. Thus, the twist is much stronger for T$_1$, where the C$_3$H$_7$ is situated close to the acceptor, than for T$_6$, where the chain is bound to the external site [Figure~\ref{geometries.fig}e)]. The pronounced torsion is a consequence of attractive N$\cdots$CH interactions between the F4-TCNQ and the alkyl chain. Longer chains like C$_3$H$_7$ can close in on the acceptor molecule and reduce the CH-N distance to 2.6~\AA, which is the corresponding equilibrium value~\cite{jackson2013}. Hence, the bonding site of the alkyl chain [position 3 versus 3' within the rings, see Figure \ref{geometries.fig}a) and e)] selectively affects the interaction with the dopant. The presence of direct coupling between alkyl chains and doping molecules demands the explicit inclusion of longer chains in the simulations of CT complex in order to obtain an accurate description of the system. However, already attached methyl groups provide a good estimate of the overall trend.

We now proceed to analyzing the electronic properties of the CT complex. Its frontier orbitals are dominated by the hybridization between the HOMO of the donor and the LUMO of the acceptor, which give rise to an occupied bonding and an unoccupied antibonding orbital superposition~\cite{zhu,ana1,ana2,thomas2019}. Also other valence orbitals show signs of hybridization, \textit{i.e.} they are delocalized over the whole complex and occur in bonding-antibonding pairs\cite{zhu}. As we will discuss in Section \ref{optical_properties_ctc.sec} in the context of optical excitations, this does not necessarily mean that the electron densities associated with these hybrid orbitals are equally distributed between the two constituent molecules.

The natural population analysis\cite{npa} allows us to pinpoint the interaction-induced charge relocalization associated with the hybridization. The ground-state CT between the H-terminated 6T and F4-TCNQ amounts to 0.28~$e$ \textit{in vacuo}. Upon closer inspection of the charge distribution within the 6T [Table \ref{resolvedCharges.tbl}a)], it is evident that the outer thiophene rings, T$_1$ and T$_6$, give only minor contributions to the charge transferred to the F4-TCNQ. The electron depletion is thus mainly restricted to the four T rings directly underneath the acceptor, as observed experimentally~\cite{localizedCT}. The corresponding positive charge is uniformly distributed among those four rings.

The addition of alkyl chains of increasing length to 6T leads to a systematic enhancement of the CT in the ground state (see Figure~\ref{ct.fig}; values specified in Section~S2 of the Supplementary Material). Upon inclusion of CH$_3$ groups, the CT goes up to 0.38~$e$; extending the chain length increases this value, saturating at 0.45~$e$. This result is consistent with the behavior of the IP of 6T, which decreases upon addition of alkyl groups [Figure~\ref{levels.fig}b)]: The lower the IP with respect to the electron affinity of the acceptor, the stronger the CT driving force~\cite{bender1986}. We recall that the reduction of the IP is mainly a direct consequence of the presence of the alkyl chains, although the chain-length dependent decrease of the torsion underneath the F4-TCNQ [Table~\ref{torsion.tbl}a)] also raises $E_{\mathrm{HOMO}}$. In contrast to the uniform distribution discussed previously, upon alkylization the excess positive charge is increasingly localized on the center rings, T$_3$ and T$_4$ [Table~\ref{resolvedCharges.tbl}a)]. Furthermore, the CT complex acquires a dipole moment of 1.2~--~1.6~D (depending on the alkyl chain length) in the $xy$-plane [see coordinate system in Figure~\ref{geometries.fig}e)], due to the breaking of its $C_2$ symmetry upon alkyl substitution. This dipole moment, induced by \textit{intra}molecular CT, is non-negligible compared to the interfacial $z$-directed dipole moment of 3.3~-~3.6~D, which is related to the \textit{inter}molecular CT. The emergence of a dipole moment parallel to the 6T axis can be rationalized by appreciating the asymmetric distances between the electronegative atoms (N and F) of the acceptor and the alkyl chains of 6T on the left- and right-hand sides of the complex, which cause an overall in-plane charge imbalance. 

\begin{table}[]
\caption{Partial charges in the donor of the charge-transfer complex in the ground state (in $e$) on the outer (T$_1$ and T$_6$), intermediate (T$_2$ and T$_5$), and inner (T$_3$ and T$_4$) thiophene rings [see Figure \ref{geometries.fig}a)] (a) \textit{in vacuo} with different alkyl groups, (b) in different solvents (increasing polarity), with CH$_3$ groups attached}
\begin{flushleft}
\footnotesize{a)}
\end{flushleft}
    \centering
    \begin{tabular}{p{1.2cm}p{1.2cm}p{1.2cm}p{1.2cm}p{1.2cm}}
    \hline
        \centering\textbf{group} & \centering outer  & \centering interm. & \centering inner & \centering total \arraybackslash \\
         \hline\centering none & \centering 0.05 & \centering 0.10 & \centering 0.12 & \centering 0.28 \arraybackslash\\
         \centering CH$_3$ & \centering 0.06 & \centering 0.13 & \centering 0.20 & \centering 0.39\arraybackslash\\
         \centering C$_2$H$_5$ & \centering 0.06 & \centering 0.15 & \centering 0.24 & \centering 0.44\arraybackslash\\
         \centering C$_3$H$_7$ & \centering 0.05 & \centering 0.15 & \centering 0.25 & \centering 0.45\arraybackslash\\
         \hline
    \end{tabular}

    \begin{flushleft}
\footnotesize{b)}
    \end{flushleft}
    \centering
    \begin{tabular}{p{1.2cm}p{1.2cm}p{1.2cm}p{1.2cm}p{1.2cm}}
    \hline
        \centering\textbf{solvent}  & \centering outer & \centering interm. & \centering inner & \centering total \arraybackslash\\
         \hline\centering none  & \centering 0.06 & \centering 0.13 & \centering 0.20 & \centering 0.39 \arraybackslash\\
         \centering C$_6$H$_6$  & \centering 0.06 & \centering 0.14 & \centering 0.24 & \centering 0.45 \arraybackslash\\
         \centering CHCl$_3$  & \centering 0.07 & \centering 0.14 & \centering 0.27 & \centering 0.48 \arraybackslash\\
         \centering CH$_3$NO$_2$  & \centering 0.07 & \centering 0.14 & \centering 0.31 & \centering 0.52 \arraybackslash\\
         \hline
    \end{tabular}
\label{resolvedCharges.tbl}
\end{table}

Solvents as well increase significantly the ground-state CT in the complex, with the degree of solvent polarity playing a crucial role (Figure~\ref{ct.fig}). In the H-passivated 6T, the ground-state CT increases from 0.28~$e$ \textit{in vacuo} to 0.36~$e$ in CH$_3$NO$_2$. With covalently bonded CH$_3$ groups, the value grows from 0.39~$e$ to 0.52~$e$, corresponding to an increase of about 30\% in both cases.
The enhanced CT is straightforwardly rationalized with electrostatic considerations. As mentioned before, the complex has a dipole moment in the stacking direction due to intermolecular CT. The reaction field due to the solvent polarization is oriented such that it increases the dipole moment of the system by additionally polarizing it, thereby enhancing the CT. As such, the solvent polarization charges cause an additional driving force for CT. As the strength of the reaction field is roughly proportional to the Onsager factor $2(\epsilon-1)/(2\epsilon+1)$~\cite{onsager}, we find an approximately linear increase in CT as a function of $2(\epsilon-1)/(2\epsilon+1)$, see Figure~\ref{ct.fig}. Since the factor $2(\epsilon-1)/(2\epsilon+1)$ rapidly approaches unity as a function of $\epsilon$, the enhancement effect saturates quickly. Hence, two different polar ($\epsilon>10$) solvents influence the solute in a very similar way, irrespective of the exact value of $\epsilon$, if no specific solute-solvent interactions (\textit{e.g.}, hydrogen bonding) come into play. 
%If this is the case, explicit solvent molecules should be included in the simulation to model these interactions, in addition to the continuum solvent.

Performing ground-state calculations \textit{in vacuo} with the geometries optimized in solution reveals that the increase of CT is mainly caused directly by the reaction field (Figure~\ref{ct.fig}). Changes in the geometry increase the CT only by a small amount, although this increase becomes slightly larger when alkyl groups are attached.

\begin{figure}
    \centering
    \includegraphics[width=0.48\textwidth]{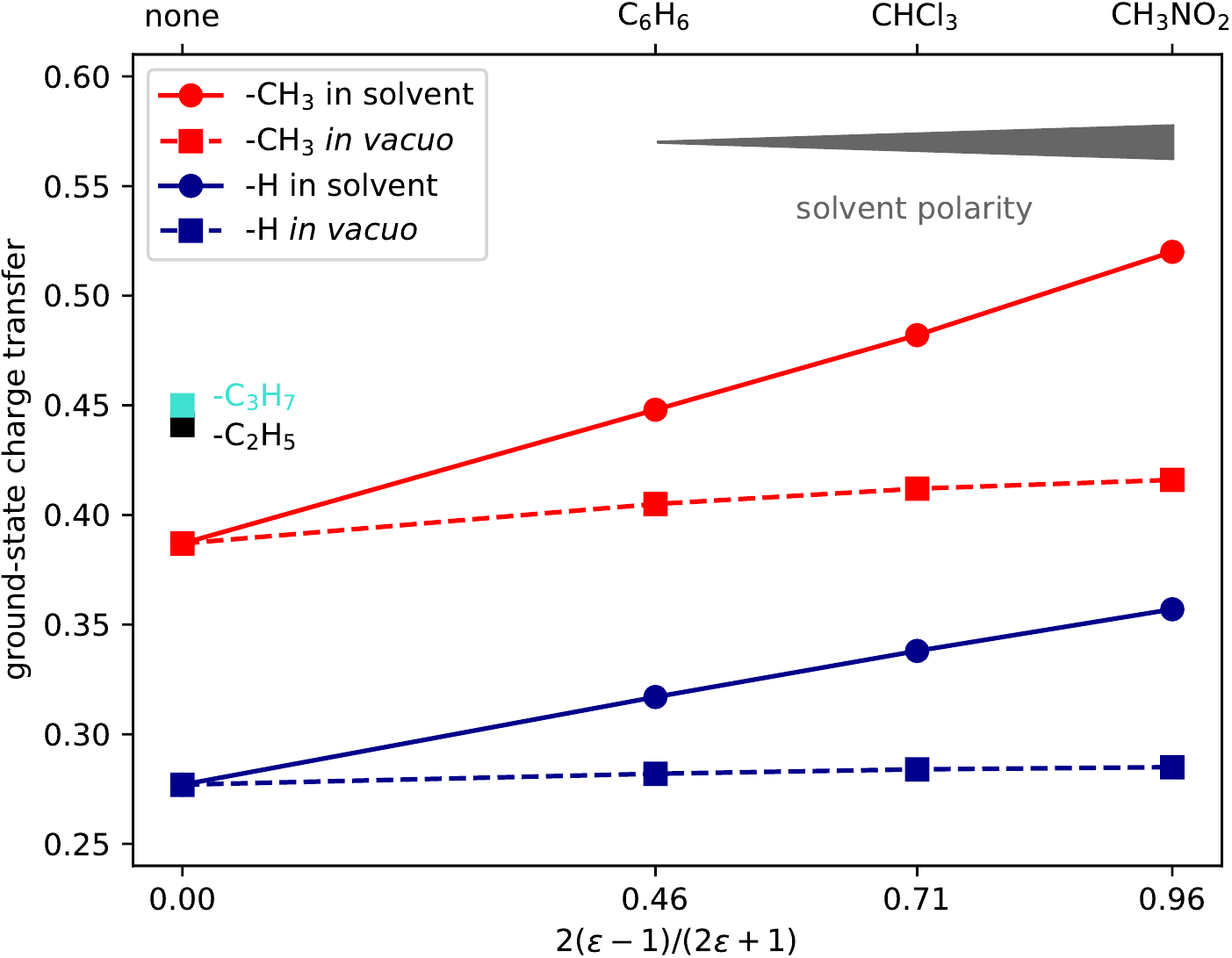}
    \caption{Ground-state charge transfer in the complex as a function of the Onsager factor $2(\epsilon-1)/(2\epsilon+1)$, where $\epsilon$ is the dielectric constant of the solvent. Circles and solid lines correspond to calculations in an implicit solvent, whereas squares and dashed curves are the results of calculations \textit{in vacuo}, using the solvent-optimized geometry. Red symbols indicate systems with CH$_3$ groups attached to the 6T, blue without alkyl chains, black and cyan the values for C$_2$H$_5$ and C$_3$H$_7$-functionalizations \textit{in vacuo}, respectively.}
    \label{ct.fig}
\end{figure}

 By resolving the partial charges on the CH$_3$-substituted 6T in the complex in different solvents [Table \ref{resolvedCharges.tbl}b)], we find the extra charge to be increasingly localized on the two inner T rings, T$_3$ and T$_4$. Larger solvent polarity only enhances the charge donation from these two monomers, further differentiating the inner rings (T$_3$ and T$_4$) from the intermediate ones (T$_2$ and T$_5$).

%%%%%%%%%%%%%%%%%%%%%%%%%%%%%
\subsection{Optical properties}\label{optical_properties.sec}
The variation of the electronic structure in the considered systems is reflected in their optical properties. We first focus on the isolated 6T in its various conformations before analyzing the (alkyl-functionalized) 6T/F4-TCNQ complex. 

%%%%%%%%%%%%%%%%%%
\subsubsection{6T \textit{in vacuo} and in solution}
The absorption spectrum of 6T is dominated by a strong peak around 3.0~eV, stemming from the HOMO-LUMO transition~\cite{tightBinding,cocc-drax15prb} (details about the orbital transitions of the first five excited states are listed in Section~S3 of the Supplementary Material).
As seen in Figure~\ref{tddft6T.fig}a), the excitation energy as a function of the backbone conformation and of the alkyl chain length follows the same trend as the electronic gap [Figure \ref{levels.fig}c)]. The oscillator strength (OS) is maximized in the most planar configurations, \textit{i.e.} when no alkyl chains are attached [Figure~\ref{tddft6T.fig}b)], and decreases with increasing chain length. We attribute this behavior to an overall reduction of the overlap between the HOMO and the LUMO upon increasing torsion angle.

\begin{figure*}
    \centering
    \includegraphics[width=\textwidth]{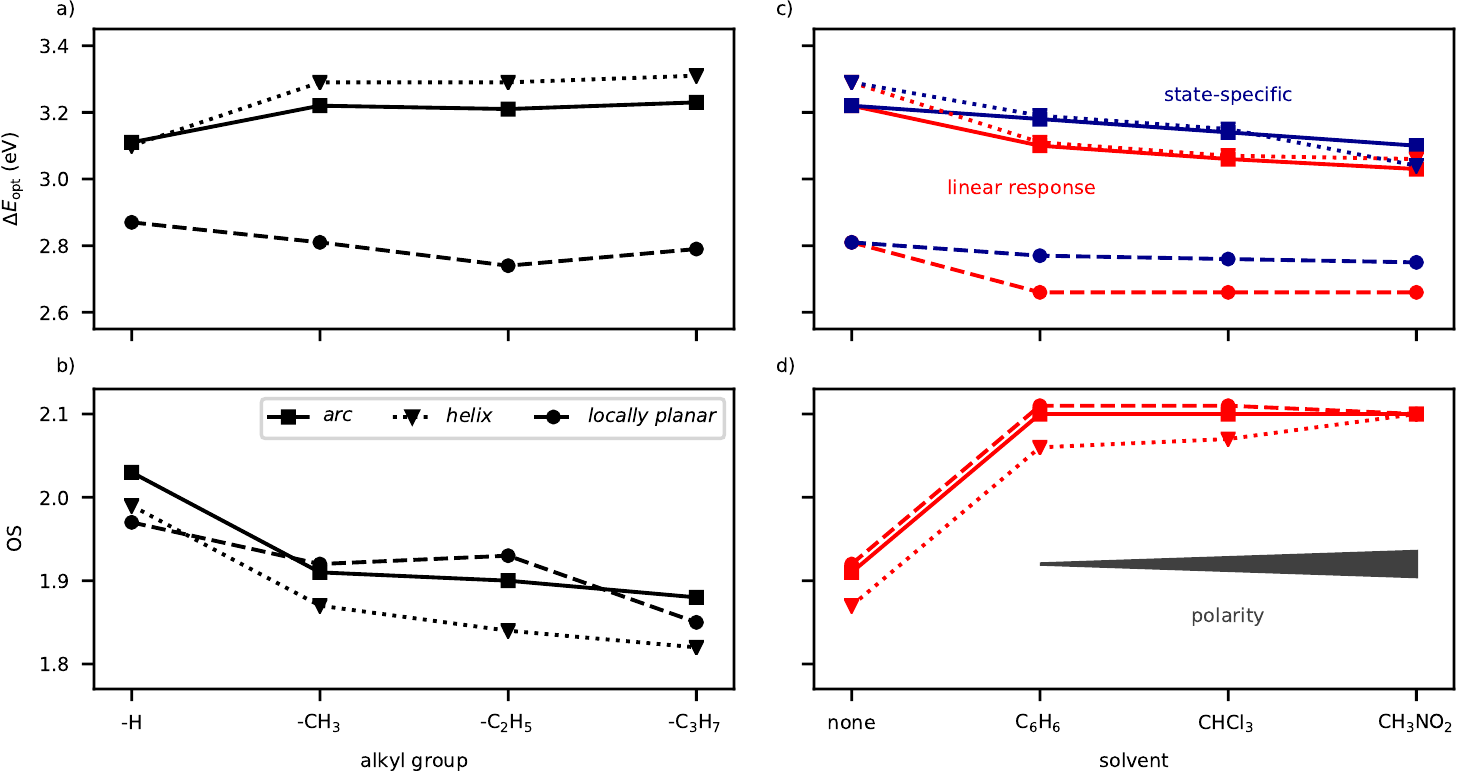}
    \caption{a) Optical gap ($\Delta E_{opt}$) and b) oscillator strengths (OS) of 6T in the inspected configurations (see Figure~\ref{geometries.fig}) \textit{in vacuo} and with alkyl chains of increasing length. c) Excitation energies, computed within linear-response TDDFT (red) and state-specific TDDFT (blue), and d) OS of CH$_3$-functionalized 6T in solution with solvents of increasing polarity.}
    \label{tddft6T.fig}
\end{figure*}

\begin{table}[]
\caption{Magnitude of the bathochromic shifts (in meV) for the HOMO-LUMO transition of the three indicated conformations of CH$_3$-functionalized 6T, and the excitations $P_1$ and $P_2$ in the 6T/F4-TCNQ CT complex, for the three considered solvents, calculated from a) linear-response TDDFT and b) state-specific TDDFT}
       \begin{flushleft}
    \footnotesize{a)}
    \end{flushleft}
    \centering
     \begin{tabular}{c p{1.5cm}p{1.5cm}p{1.5cm}}
     \hline
    \textbf{excitation} & \centering C$_6$H$_6$ & \centering CHCl$_3$ & \centering CH$_3$NO$_2$\arraybackslash\\
         \hline H$\rightarrow$L ($arc$) & \centering 120 & \centering 160 & \centering 190\arraybackslash\\
         H$\rightarrow$L ($helix$) &\centering 180 & \centering 220 & \centering 230 \arraybackslash\\
          H$\rightarrow$L ($planar$) & \centering 150 & \centering 150 & \centering 150\arraybackslash\\
          $P_1$ & \centering 50 &  \centering  40 & \centering 30\arraybackslash\\
          $P_2$ & \centering 70 &  \centering  70 & \centering 70\arraybackslash\\
         \hline
    \end{tabular}
    \label{shifts.tab}
  \begin{flushleft}
    \footnotesize{b)}
    \end{flushleft}
    \centering
\begin{tabular}{c p{1.5cm}p{1.5cm}p{1.5cm}}
\hline
          \textbf{excitation} & \centering C$_6$H$_6$ & \centering CHCl$_3$ & \centering CH$_3$NO$_2$ \arraybackslash\\\hline
          H$\rightarrow$L ($arc$) & \centering 40  & \centering  80 & \centering 120\arraybackslash\\
          H$\rightarrow$L ($helix$) & \centering 100 & \centering 140 & \centering 250\arraybackslash\\
          H$\rightarrow$L ($planar$) & \centering 40  & \centering  50 & \centering 60\arraybackslash\\
          $P_1$ & \centering 30  & \centering  10 & \centering 0\arraybackslash\\
          $P_2$ & \centering 170 & \centering 160 & \centering 150\arraybackslash\\
         \hline
    \end{tabular}
\end{table}

The photoabsorption characteristics can be influenced by solvation effects in two ways. The ground-state geometries and electronic structures are influenced by the solvent, which, in turn, affects the optical spectrum via its dielectric constant. Additional changes are directly related to the interaction between the dynamical electron density of the excited solute and the solvent, and are determined by the refractive index of the solvent. In the event of an optical excitation, the induced density of the molecule, \textit{i.e.} the difference between the excited, time-dependent electron density and the ground-state density, is given by the superposition of the transition density of the excitation and the stationary density difference of the two states involved~\cite{myFirstPaper}. Both these charge densities interact with the fast degrees of freedom of the solvent, each contributing to a total solvatochromic shift. The transition density oscillates at the transition frequency, giving rise to a corresponding in-phase solvent polarization. This represents a stabilizing induced dipole$\cdots$induced dipole interaction, generated by \textit{dispersion} forces\cite{mcrae}. The density difference, on the other hand, is static. It also polarizes the solvent and interacts with the corresponding reaction field. These interactions are of \textit{electrostatic} origin. 
It is important to make this distinction, as linear-response TDDFT captures the dispersion contribution, but only part of the electrostatic one~\cite{lrVsSs1, lrVsSs2}. Specifically, it misses the adaptation of the solvent to the density difference and only considers the interaction between the density difference and the \textit{ground-state} polarization charges. The state-specific method, on the other hand, captures the electrostatic interactions in full, including solvent relaxation, but lacks the dispersion part\cite{lrVsSs1, lrVsSs2}. Thus, the two methods are complementary in the prediction of solvatochromic shifts; in case the dispersion part is dominant, linear-response TDDFT turns out to be more accurate, whereas state-specific TDDFT is more appropriate when the electrostatic contribution is the major one. As a rule of thumb, dispersion forces are stronger for local excitations with high OS, whereas the electrostatic contribution is larger for charge-transfer excitations, which are characterized by a large density redistribution and low OS. In terms of computational costs, the linear-response method is much cheaper. It yields multiple excited states at once, alongside transition properties such as the OS. The state-specific method, on the other hand, requires two calculations for a single excited state, and yields only the corresponding excitation energy. Thus, it is usually put to use more selectively for states of particular interest~\cite{caricato2013jcp}. 

An indicator for the relative weight of the aforementioned contributions (dispersion forces \textit{vs.} electrostatic interactions) are the two dipole moments characterizing an excitation: the transition dipole moment $\boldsymbol{\mu}_{g\rightarrow e}$, related to the peak height, as the OS is proportional to its square modulus, and the static dipole difference, expressed by $\boldsymbol{\mu}_e-\boldsymbol{\mu}_g$\cite{lrVsSs1,lrVsSs2}. In the case of the HOMO-LUMO transition in 6T, $|\boldsymbol{\mu}_{g\rightarrow e}|\approx$~10~D and $|\boldsymbol{\mu}_e-\boldsymbol{\mu}_g|$~$<$~1~D, which clarifies that the bathochromic shifts are mainly due to dispersion interactions. Thus, the linear-response formalism features significantly larger shifts compared to the state-specific one [see Figure~\ref{tddft6T.fig}c) and Table~\ref{shifts.tab}]. The state-specific shift is still sizeable, as a large portion of it is related to solvation-induced structural distortions rather than direct interactions. Consequently, the overall shifts are most pronounced in the \textit{helix} configuration, as its geometry is particularly sensitive to external perturbations, as evidenced by the large differences in the torsion angles in the different solvents [Table~\ref{torsion.tbl}b)].

Regardless of the adopted flavor of PCM/TDDFT, the redshift of the HOMO-LUMO peak of 6T becomes larger for increasing solvent polarity [Figure~\ref{tddft6T.fig}c)]. This increase cannot be explained in terms of dynamical interactions between solute and solvent, as the refractive index $n$ is almost equal in all considered solvents (1.38-1.50). Hence, the solvent-induced changes to the ground-state properties and to the electronic structure, which are instead related to the dielectric constant $\epsilon$, are the underlying cause of the redshift increase.  Indeed, as noted in Section~\ref{electronic_properties_6t_sol.sec}, the solvent decreases the torsion angles in the 6T according to its polarity [see Table \ref{torsion.tbl}b)], which correspondingly narrows electronic and optical gaps. In general, the OS increases when the molecule is immersed in a solvent [Figure \ref{tddft6T.fig}d)], as a consequence of reduced torsion and direct interactions with the solvent. The rather small changes upon increasing the polarity suggest the direct interactions to be the key factor in this context. 

Finally, we inspect the optical properties of the isolated \textit{locally planar} 6T extracted from the geometry of the CT complex formed with F4-TCNQ. Functionalizations with alkyl chains of increasing length are considered. Similarly to what is observed for the \textit{arc} and \textit{helix} configurations, the computed optical gaps follows the trend obtained for the electronic gaps, exhibiting only a rigid shift with respect to them [Figure~\ref{levels.fig}c) and \ref{tddft6T.fig}a)]. We recall that these energies are generally lower than in the other configurations, as a result of the increased planarity of the four T rings interacting with the F4-TCNQ, which in turn implies increased coupling between the monomers. This effect is partly counteracted by the increased torsion of T$_1$ and T$_6$, which is also responsible for the anti-trend gap opening observed in the C$_3$H$_7$-functionalized 6T. The corresponding sharp drop in the OS [see Figure~\ref{tddft6T.fig}b)] is a consequence of the enhanced localization of the HOMO and the LUMO on the center of the molecular backbone, overall diminishing the orbital density localized on T$_1$, which is weakly coupled to the other rings. This phenomenon leads to a decrease of the magnitude of the transition dipole moment, which can be approximately expressed as 
\begin{align}
    \boldsymbol{\mu}_{\mathrm{H}\rightarrow\mathrm{L}} \approx \int\text d^3r\,\psi^*_{\mathrm{LUMO}}(\textbf{r})\,(-\textbf{r})\,\psi_{\mathrm{HOMO}}(\textbf{r}).
    \label{eq:mu-HL}
\end{align}
due to the dominant HOMO $\rightarrow$ LUMO character of the excitation ($\sim$90\%, see Section~S3 in the Supplementary Material).
It is evident from Eq.~\eqref{eq:mu-HL} that $\boldsymbol{\mu}_{\mathrm{H}\rightarrow\mathrm{L}}$ depends on the overlap as well as on the spatial extent of the orbitals, due to the presence of the dipole operator, $-\textbf{r}$. The OS is proportional to $|\boldsymbol{\mu}_{\mathrm{H}\rightarrow\mathrm{L}}|^2$.

Moving on to the solvation effects, we find that the CH$_3$-functionalized, \textit{locally planar} 6T experiences a bathochromic shift in solution [Figure \ref{tddft6T.fig}c)]. Contrary to the \textit{arc} and \textit{helix} structures, this shift is independent of the solvent polarity. For the undoped structures, we rationalized the dependence of the excitation energy on the solvent polarity in terms of changes of the underlying geometry. The geometry of the CT complex, however, is much less affected by the solvent, since the coupling between the 6T and the F4-TCNQ is stronger than the electrostatic interactions with the solvent molecules, and, as such, defines the local structure of the complex. 

%%%%%%%%%%%%%%%%%%%%%
\subsubsection{6T/F4-TCNQ charge-transfer complex \textit{in vacuo} and in solution}\label{optical_properties_ctc.sec}
\begin{figure*}
    \centering
    \includegraphics[width=\textwidth]{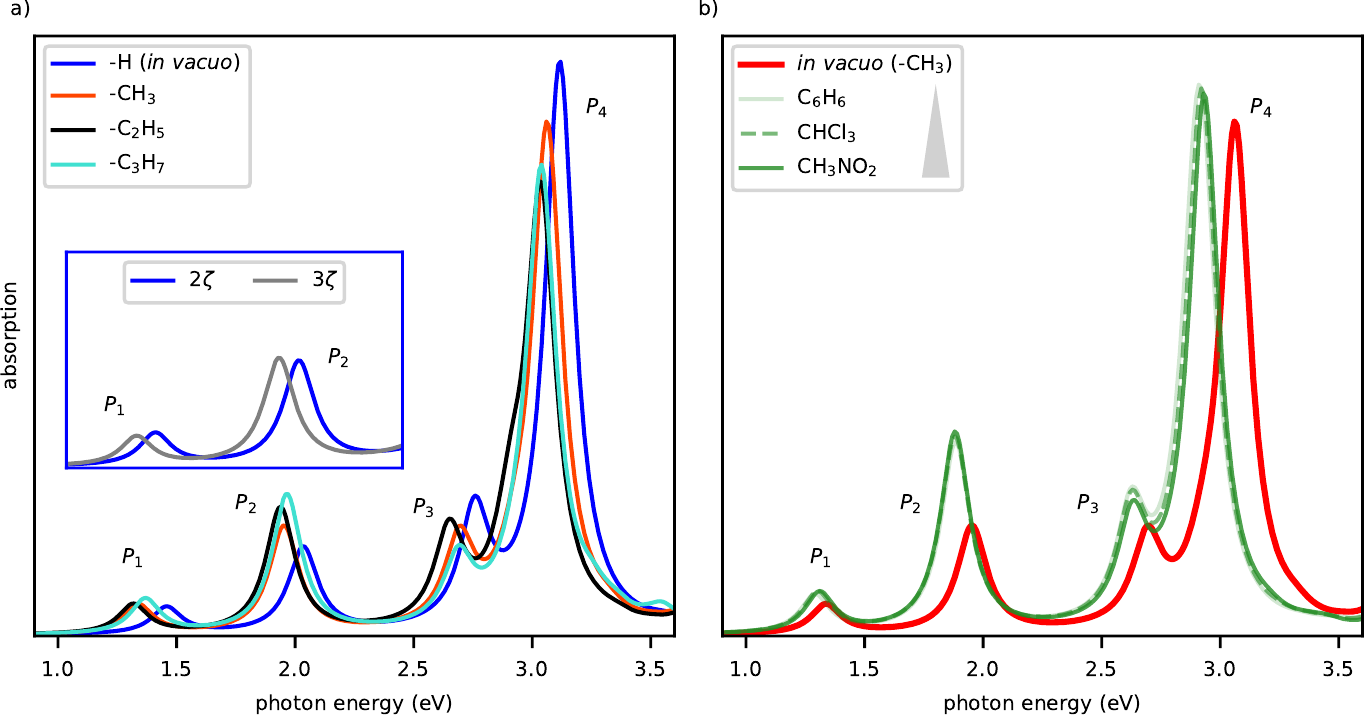}
    \caption{Optical absorption spectra of the CT complexes, calculated with linear-response TDDFT: a) \textit{in vacuo} with alkyl chains of increasing length, and b) in different solvents, with CH$_3$ groups attached to the 6T. The inset in panel a) shows the first two peaks of the spectrum of the H-terminated complex, as calculated with the cc-pVDZ (2$\zeta$) and cc-pVTZ (3$\zeta$) basis sets.}
    \label{tddftCpx.fig}
\end{figure*}

The previous discussion leads us to the analysis of excited states of the 6T/F4-TCNQ complex. We start by considering the system with the H-terminated 6T as a donor. In this case, the linear absorption spectrum [Figure~\ref{tddftCpx.fig}a)] shows two features at low energy, $P_1$ and $P_2$, at 1.5~eV and 2.0~eV, respectively, which cannot be assigned to either molecular component individually. Indeed, they correspond to excitations involving the hybridised frontier states (hereafter called \textit{hybrid excitations}). $P_1$ is formed by a transition from the HOMO to the LUMO, and $P_2$ by a transition from the HOMO-1 to the LUMO of the complex~\cite{zhu, ana1, ana2}, respectively (details about the first ten excited states and their constituting orbital transitions are listed in Section~S4 of the Supplementary Material). They are both experimentally detectable signatures of CT complex formation~\cite{mendez2013}. At higher energies, the maximum $P_3$ corresponds likewise to a hybrid excitation, whereas the strongest peak at 3.0~eV, $P_4$ [see Figure~\ref{tddftCpx.fig}a)], is related to the intramolecular HOMO-LUMO transitions of the individual constituents. 

The accuracy of excited-state calculations is known to be improved by adding diffuse functions to the basis set.~\cite{elliott+2008rcc} We investigate their effect by comparing results obtained with the 6-31G(d,p), 6-31+G(d,p), and 6-31++G(d,p) basis sets, which feature no diffuse functions, diffuse functions on heavy atoms, and diffuse functions on heavy atoms and hydrogens, respectively (all results in Section S4.5 in the Supplementary Material). Employing 6-31+G(d,p) yields excitations that are systematically redshifted by about 50~meV with respect to those obtained with 6-31G(d,p), whereas going one step further to 6-31++G(d,p) does not lead to additional improvements. The OS predicted by these three basis sets differ by 0.03 at maximum. Comparing the double-$\zeta$ cc-pVDZ and triple-$\zeta$ cc-pVTZ basis sets, we find similarly small improvements in the spectra [see inset of Fig.~\ref{tddftCpx.fig}a) and Section S4.4 in the Supplementary Material].
In summary, enlarging the basis set leads to only small and predictable improvements of the spectra, while at the same time it increases significantly the computational complexity. We conclude that the double-$\zeta$ basis set without diffuse functions is sufficient for our purposes and allows us to deal efficiently also with the largest of the complexes.

By means of natural population analysis of the excited states calculated via state-specific TDDFT, we find that in both hybrid excitations, $P_1$ and $P_2$, the CT is enhanced with respect to the ground state. As discussed in Section~\ref{electronic_properties_ctc.sec}, the ground-state electron transfer from the 6T to the F4-TCNQ \textit{in vacuo} amounts to 0.28~$e$. In the first excited state, corresponding to $P_1$, it is increased to 0.60~$e$. This behavior can be understood by analyzing the character of the HOMO and the LUMO of the complex, which are involved in this transition. While they correspond to bonding and antibonding superpositions of the HOMO of the 6T and of the LUMO of the F4-TCNQ, the HOMO of the complex has predominantly the character of the HOMO of the 6T, while the LUMO of the complex is more resemblant of the LUMO of the F4-TCNQ~\cite{thomas2019}. This is a general feature of CT complexes; hence, the excited state corresponding to the antibonding superposition has a more ionic character~\cite{bender1986}. In the excited state $P_2$, the CT is further increased to 0.78~$e$, thus approaching an integer value. This is a consequence of the initial state of this transition, the HOMO-1 of the complex, being mainly localized on the 6T, in spite of being a hybridized orbital~\cite{ana1}. The excited states $P_3$ and $P_4$ lead to CTs of 0.29~$e$ and 0.36~$e$, respectively.

The addition of alkyl chains to the F4-TCNQ-doped 6T causes an overall red-shift of the absorption spectrum of the complex [Figure~\ref{tddftCpx.fig}a)].
This characteristic can be traced back to the readjustment of the orbital energies of 6T, due to the dopant-induced structural distortions and to the reduction of the IP induced by the alkyl groups (see above). In particular, the shift of $P_4$, which corresponds to the two intramolecular HOMO-LUMO transitions in the individual constituents, is related to structural changes in the alkylated 6T. Indeed, the shift can be observed already in the \textit{locally planar} 6T, and is related to the increased planarization of the four T rings underneath the acceptor upon prolongation of the alkyl chains [see Table~\ref{torsion.tbl}a)]. The redshifts of the hybrid peaks, $P_1$ and $P_2$, are a consequence of the generally increased energies of the orbitals of 6T due to alkyl functionalization, which also raises the energies of the hybrid orbitals of the complex. The LUMO of the complex, though, corresponds mainly to the LUMO of the F4-TCNQ, as discussed above, and is thus less affected by the level readjustment in the 6T. As a consequence, its energy does not shift up as much as that of the HOMO, giving rise to reduced electronic and optical gaps in the alkylized complexes. Hence, the spectral shifts displayed in Figure~\ref{tddftCpx.fig}a) have different causes depending on the character of the excitation: The shift of $P_4$ is related to the planarization-induced bandgap decrease of 6T, whereas the shift of the hybrid excitations is related to the direct electronic influence of the alkyl chains. The small anti-trend blue-shift of $P_1$ and $P_2$ upon increasing the alkyl chain length from C$_2$H$_5$ to C$_3$H$_7$ is caused once again by the effect of the pronounced torsion of T$_1$, due to attractive interaction between the C$_3$H$_7$ chain and the N atoms in the F4-TCNQ, which decreases the energy of the highest occupied orbitals of the 6T.

Solvents cause an overall redshift of the absorption [Figure~\ref{tddftCpx.fig}b)], ranging from 50~meV ($P_1$) to 150~meV ($P_4$). In the following, we focus mainly on the hybrid excitations, $P_1$ and $P_2$, which represent a fingerprint of CT complexes. Since both excitations cause additional CT from the 6T to the F4-TCNQ (see above), they increase the static dipole moment of the complex in the $z$-direction [see coordinate system in Figure~\ref{geometries.fig}e)]. Thus, they offer the opportunity to compare linear-response and state-specific TDDFT results. In the case of $P_1$, we find a transition dipole moment of $|\boldsymbol{\mu}_{g\rightarrow e}|\approx$~5~D and a static dipole difference of $|\boldsymbol{\mu}_{e}-\boldsymbol{\mu}_{g}|\approx$~2~D, which is consistent with the relative shifts from linear-response and state-specific methods (Table~\ref{shifts.tab}). In the case of $P_2$, on the other hand, $|\boldsymbol{\mu}_{g\rightarrow e}|\approx$~7~D and $|\boldsymbol{\mu}_{e}-\boldsymbol{\mu}_{g}|\approx$~4~D, but the state-specific shift is much larger than the linear-response one (see Table~\ref{shifts.tab}). Hence, the relative magnitudes of the two dipole moments do not necessarily predict the relative shifts quantitatively. However, we can clearly see how a large change of the static dipole moment entails a significant redshift which is captured only by state-specific TDDFT. This method is therefore more appropriate for excitations with CT character such as $P_2$. Notably, $P_2$ is still a moderate representative of CT excitations; long-range CT excitations with diminished orbital overlap can have an OS close to zero (and thus, $|\boldsymbol{\mu}_{g\rightarrow e}|\approx0$), but a static dipole difference of order $\sim$10~D.

\begin{table}[]
\caption{Charge transfer (in $e$) from methylized 6T to F4-TCNQ in the ground state and in the excited states $P_1$ and $P_2$}
    \centering
    \begin{tabular}{p{1.4cm}p{1.2cm}p{1.2cm}p{1.2cm}}
    \hline
        \centering \textbf{solvent}  & \centering GS & \centering $P_1$ & \centering $P_2$\arraybackslash \\
         \hline \centering none  & \centering 0.39 & \centering 0.60 & \centering 0.78 \arraybackslash\\
          \centering C$_6$H$_6$ & \centering 0.45 & \centering 0.68 & \centering 0.88 \arraybackslash\\
          \centering CHCl$_3$  & \centering 0.48 & \centering 0.73 & \centering 0.95 \arraybackslash \\
          \centering CH$_3$NO$_2$  & \centering 0.52 & \centering 0.80 & \centering 1.04 \arraybackslash\\
         \hline
    \end{tabular}
    \label{ct_excited_states.tab}
\end{table}

The presence of a solvent additionally increases the OS of all excitations. Particularly $P_2$ grows notably in intensity, similarly to the behavior experienced upon the addition of alkyl chains [Figure \ref{tddftCpx.fig}a)]. Since the 6T of the analyzed solvated complex is methylized, we find that the effects of alkylization and solvation add up. It was previously found that the OS of $P_2$ is furthermore increased by prolonging the length of the oligothiophene backbone, which concomitantly causes a redshift of the peak energy~\cite{ana1}. Combined with the fact that the redshift seen in Figure~\ref{tddftCpx.fig}b) is already underestimated by the linear-response formalism, this result suggests that in the polymer limit, $P_2$ might come energetically close to $P_1$ or even surpass it, while having much higher absorption strength. Differences between the spectra in different solvents are barely noticeable, which is a consequence of the increased stiffness of the overall structure. While the twist angles in the 6T alone represent loose degrees of freedom with shallow potential energy curves\cite{gierschner2007}, the torsion is locked in the vicinity of F4-TCNQ. Thus, the properties of the former are more strongly correlated with the parameters of the environment.

We evaluate the excited-state CT from the alkylated 6T to the F4-TCNQ in solution by conducting a natural population analysis on the density obtained with non-equilibrium state-specific TDDFT.  Similarly to the ground-state case, we obtain that the CT in the excited states $P_1$ and $P_2$ is enhanced by the solvent polarization (Table~\ref{ct_excited_states.tab}). Notably, a higher solvent polarity leads to a larger increase in CT, despite the refractive indices of the different solvents being similar. The polarity-dependent enhancement is thus related to the part of the ground-state reaction field that stays frozen in the non-equilibrium calculation. In the excited state $P_2$, the CT is higher than 1.

We recall that the scenario described above corresponds to \textit{vertical} excited states and, as such, differs from the one of \textit{adiabatic} excited states. The latter lie at the minimum of the excited-state potential energy surface and are reached after internal nuclear relaxation and the reorientation of solvent molecules upon photo-absorption. This scenario can be explored after an excited-state geometry optimization with linear-response TDDFT and equilibrium PCM, also giving access to fluorescence wavelengths~\cite{furche2002jcp, scalmani2006jcp}. Such an analysis, however, goes beyond the scope of this work.

%%%%%%%%%%%%%%%%%%%%%%%%%
\subsection{Summary and Conclusions}
We performed a comprehensive first-principles analysis of the structural, electronic, and optical properties of sexithiophene, a representative organic semiconductor molecule, in order to unravel the non-trivial interplay of all the degrees of freedom involved. For this purpose, we considered 6T \textit{in vacuo} and in solutions of increasing polarity, and we also inspected its $p$-doped and alkyl-functionalized counterparts.
We found that both solvent and alkyl chains heavily affect the structure of the molecule, by regulating the torsion angle between adjacent monomers.
In turn, this conformational variability crucially affects the electronic properties of the molecule, including energy gap, ionization potential, and band widths. Furthermore, the delocalization of valence electronic states into the alkyl chains directly increases their energy, with notable consequences for the level alignment with other molecules. The main effect of the solvent is to decrease the torsion between the thiophene rings, which in turns influences the electronic and the optical properties. Dynamical interactions between the photo-excited molecule and the solvent cause a redshift of the main absorption peak, together with an increase of its oscillator strength. 

Doping 6T with a strong acceptor like F4-TCNQ significantly impairs the flexibility of the oligomer. The formation of a charge-transfer complex planarizes the 6T backbone and locks the torsion between the rings, thereby enhancing the rigidity of the molecular structure. Hence, the torsion overall plays a less important role in this case. On the other hand, the alkylization of 6T increases the CT within the complex. Solvation leads to a further increment of CT, as the interfacial dipole moment causes a self-reinforcing electric field by polarizing the surrounding solvent molecules. The effects of alkylization and solvation on the optical properties of the 6T/F4-TCNQ complex are more diverse than in the case of the pristine 6T. The energy of the \textit{hybrid} excitations, which dominate the absorption onset, is affected by alkyl-functionalization through the variation of the energy levels of 6T, which alter the level alignment with the dopant. This, in turn, influences orbital hybridization and, consequently, the optical absorption. Peaks in the higher-energy region of the spectra, corresponding to intramolecular excitations, exhibit a similar behavior as in the pristine molecule. Comparing linear-response and state-specific TDDFT methods, we find solvatochromic shifts that are related to both dynamical and electrostatic complex-solvent interactions, with the latter being dominant in hybrid excitations with marked charge-transfer character. This type of excitations is overall most affected by alkylization and solvation, both in terms of energy and oscillator strength. 

Our results provide understanding into the correlation between solvation, alkylization, and doping, that often coexist in experimental samples of organic semiconductors.
Furthermore, our findings offer indications to simulate similar systems from first principles accounting for all their degrees of freedom in a accurate and yet efficient manner.
For example, we find that the addition of methyl groups is sufficient for modelling alkylated 6T but does not capture all the desired effects in charge-transfer complexes, where outer segments of longer alkyl chains directly couple to the acceptor.
As most of our results depend on rather general properties of organic semiconductors, such as their structural flexibility and the tunability of their electronic and optical properties upon functionalization and doping, we anticipate that this work will provide useful indications to analyze and understand computational results on this class of materials.

%%%%%%%%%%%%%%%%%%%%%%%%%%
\section{Acknowledgements}
We are thankful to Ahmed E. Mansour for useful discussions. This work was funded by the Deutsche Forschungsgemeinschaft (DFG, German Research Foundation) - project number 182087777 - SFB 951 and 286798544 – HE 5866/2-1 (FoMEDOS), by the German Federal Ministry of Education and Research (Professorinnenprogramm III) as well as by the State of Lower Saxony (Professorinnen für Niedersachsen). Computational resources are provided by the North-German Supercomputing Alliance (HLRN), project bep00076.

%\bibliography{bib}
\bibliographystyle{rsc}

\providecommand*{\mcitethebibliography}{\thebibliography}
\csname @ifundefined\endcsname{endmcitethebibliography}
{\let\endmcitethebibliography\endthebibliography}{}

\end{document}